# A Reference Architecture for Blockchain-based Peer-to-Peer IoT Applications

Gowri Sankar Ramachandran, University of Southern California, Los Angeles, USA.
Bhaskar Krishnamachari, University of Southern California, Los Angeles, USA.

## Abstract

The advent of Blockchain and Distributed Ledger Technologies enable IoT and smart city application developers to conceive new types of applications and solutions for identity management, trust, and data monetization. However, architecting blockchain-based IoT applications remain challenging due to the heterogeneous nature of blockchain platforms and lack of guidelines on how to interface existing components in the IoT ecosystem with the emerging Blockchain technology. This article explains the characteristics of blockchain and IoT technologies and presents a general reference architecture that can be used to develop many blockchain-based peer-to-peer IoT applications.

## 1. Introduction

The decentralization provided by blockchain and decentralized ledger technology enables application developers in the area of IoT and smart cities to apply this technology, especially in multi-stakeholder deployments. Real-world IoT deployments are being scaled up and starting to rely on software and hardware infrastructure owned and managed by multiple organizations and stakeholders. For example, ThingsNetwork, a non-profit organization, manages a number of LoRaWAN gateways throughout the world by employing a centralized cloud-based backend solution, through which the packets from LoRaWAN client devices flow to the application developers. As IoT deployments start to scale to city-wide including through data marketplaces and subscription-based communication model, devices will have to rely on hardware and software owned and managed by different stakeholders.

In multi-stakeholder deployments, there is a need to guarantee trust, security, and privacy to encourage the IoT device owners and application developers to engage with third-party IoT service and data providers. Besides, the introduction of data marketplaces [1] allows the device owners and application developers to build new types of applications. Thus, the future IoT applications and deployments will increasingly rely on various third-party services for communication, processing, storage, and computation while earning a financial incentive for sharing their data with other application developers.

In this article, we propose a reference architecture explaining how large-scale IoT deployments involving multiple stakeholders and data monetization can employ blockchain and distributed ledger technology to address micro-payment and trust issues. In particular, we identify four fundamental building blocks of a peer-to-peer blockchain-based IoT application that form a key part of the architecture: payment, record, identity, and control. We show that a large class of blockchain-based IoT applications can be developed using our reference architecture by presenting three example solutions focusing on payment protocol and data marketplace.

## 2. Characteristics of IoT

Cities and industries around the world are starting to adopt Internet-of-Things (IoT) technology to automate the management processes through the data provided by the sensors and to monitor and control digital devices remotely. The key functional elements of IoT include sensing, computation, communication, and control or actuation. These functionalities are realized through a combination of embedded devices, wireless communication technologies, sensors, and actuators. A careful configuration and distribution of hardware, software and networking components are essential to delivering the desired application goals. Figure 1 shows the architecture of IoT involving the hardware, software, networking along with the members in the end-to-end IoT ecosystem.

"Connected things" layer consists of sensors, actuators, embedded devices, mobile phones, and smart appliances capable of collecting data from their operational environment or actuation of electronic appliances including digital door lock, lamps, garage doors, among other things. To develop meaningful applications or services using the "connected things", it is essential to deliver the data to the desired application end-points. Depending on the application requirement, the data from the end-devices flow either to the edge layer or cloud layer, or in some cases, the edge layer aggregates the data and reports it to the cloud layer for further processing, analytics, and visualization. The communication layer enables the devices to transmit to and receive from other devices in the application stack. IoT application developers build visualization, analytics, and other applications on cloud infrastructure or mobile devices and servers. Almost all real-world IoT deployments use the architecture presented in Figure 1.

**Ecosystem Members and Their Role in IoT Deployments:** At each layer in the IoT architecture, various device owners, service providers, network operators, data consumers, and application developers are either producing data, delivering communication support, or consuming data or services. For all these activities, the members in the IoT ecosystem are owning and managing hardware and software infrastructures. In a small-scale IoT deployment, a single organization can deploy and manage the entire infrastructure, but it may not scale when the deployment consists of hundreds of devices as envisaged in smart city deployments, or the deployment of a technology especially for wireless communication requires special permission or licensing. For example, the growing interest in long-range communication technology for IoT results in a large number of IoT deployments using LoRaWAN, NB-IoT, and SigFox [1].

However, the deployment of a communication gateway for these technologies require access to a tall infrastructure to achieve better range [2], or only the licensed service providers are allowed to provide communication services to customers and device owners based on subscription. In such scenarios, device owners are required to either buy a subscription or register their devices on the central communication portal to get free access as followed by Things Network for LoRaWAN technology. Besides, the use of edge and cloud infrastructure for data processing, analytics, and other services require the involvement of various edge and cloud providers who deliver software as a service (SaaS) and platform as a service (PaaS) solutions. Such deployment models show the involvement of various stakeholders in IoT applications.

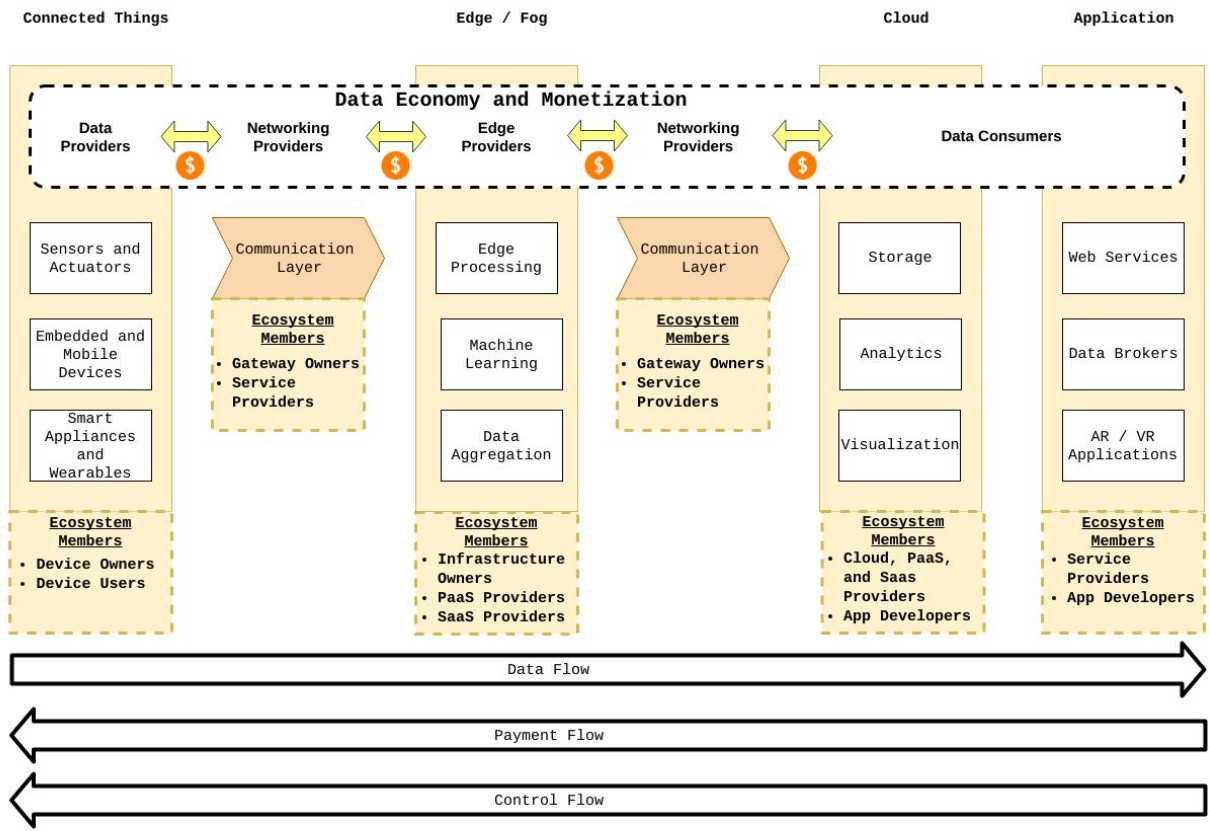

Figure 1: Architecture of IoT showing the ecosystem members and a view of the emerging data economy.

**Data Economy and Monetization:** The data produced by the IoT devices and appliances at the "things" layer adds value to a number of applications. Machine learning and AI algorithms can use the environmental sensor data to better understand the climate change or correlate the environmental data with the traffic sensor data to study the patterns and recommend routes that have less pollution. All these types of applications require data from a large collection of sensors from wide geographical neighbourhood. But, the deployment of sensors and the necessary communication infrastructure at city-scale require significant financial resource while introducing high management and maintenance complexity. "Data economy and monetization" revolve around the concept of incentivizing the device owners and mobile phone users to share their data with application developers. This new type of "data-for-value" applications are starting to emerge at the IoT and smart cities landscape through data marketplaces [3], which allow the device owners to stream their data in real-time to the applications managed by the buyers. This model not only reduce the deployment complexity but also enable the IoT technology to scale better.

*The involvement of multiple stakeholders and the emerging data economy allow the IoT applications to scale, but they require solutions for trust, payment, and identity management.*

**A Note on IoT Device Classes:** Hardware devices are central to IoT applications. Data processing, storage, Input/Output (IO) interfacing, and in some cases, the communication capabilities of devices depend on the onboard resources. The Internet Engineering Task Force's (IEFT) draft on "Terminology for Constrained-Node Networks" classifies the IoT devices into three different classes based on the program (flash) and code (RAM) memory sizes [4]: Class 0 devices are severely constrained with extremely limited resources for processing and storage. The lack of resources makes these devices less suitable for computation-intensive applications such as wireless security protocols and encryption that require significant computation and storage resources for encryption and decryption operations. Besides, class-0 devices do not connect directly to the Internet due to their limited radio capabilities, and in all cases, the devices in this category rely on a gateway or a proxy server for Internet connectivity. Class 1 devices have limited constraints, and they have enough computation and storage resources for running a lightweight network stack. Devices in this category are capable of supporting messaging protocols such as CoAP [5] and MQTT [6] besides enabling IP-based communication. Lastly, Class 2 devices are less resource constrained and are capable of running a complete network stack similar to the one used in notebooks and laptops. Typically, class-2 devices are used as a communication gateway for enabling Internet-connectivity to other resource-constrained devices. Blockchain-based IoT applications have to be architected taking into account the device classes since it determines their processing capabilities.

In the next section, we will describe the characteristics of blockchain and distributed ledger technologies and explain how these technologies can be used for multi-stakeholder IoT deployments and to provide support for emerging IoT-driven data economy.

# 3. Introduction and Overview of Blockchain and Distributed Ledger

The introduction of Bitcoin paved the way for a novel and decentralized payment system with built-in mechanism for trust. Notably, the Bitcoin system removed the need for a centralized entity to manage and verify the payment transactions. Ever since the arrival of Bitcoin, a number of platforms have been developed to provide support for peer-to-peer transactions in a decentralized setup. The Blockchain technology provides a combination of cryptography, consensus algorithm, distributed ledger and, optionally, decentralized computation capability, to create a decentralized and trustworthy platform.

**Cryptographic Digital Signature:** The public-key cryptography is used in blockchain to generate a signature for Blockchain transactions. Users carry out transactions by creating a digital signature using their private keys. Recipients in the blockchain network verify the transaction using the public key of the sender to ensure that the transaction is indeed signed by the sender. Source or end-devices sign the transactions when they create a transaction.

**Distributed Ledger:** Blockchain use a distributed storage to record the transactions. In essence, all the platforms in the network store either all the transactions or a subset of transactions. All the nodes in the network come to a consensus (using a consensus algorithm) before entering the transactions into the ledger. This feature makes blockchain effectively immutable. While initially blockchain protocols were designed for cryptocurrency transactions, in general, any data could be included in the ledger.

**Consensus algorithm:** Blockchain does not rely on a centralized server for verification and validation of transactions. Instead, Blockchain uses a peer-to-peer model, and all the decisions within the network are made by the participating members through a consensus protocol. There are many different consensus approaches ranging from permissionless proof of work with longest-chain adoption in Bitcoin to the permissioned Byzantine fault-tolerant protocol used in Tendermint [7].

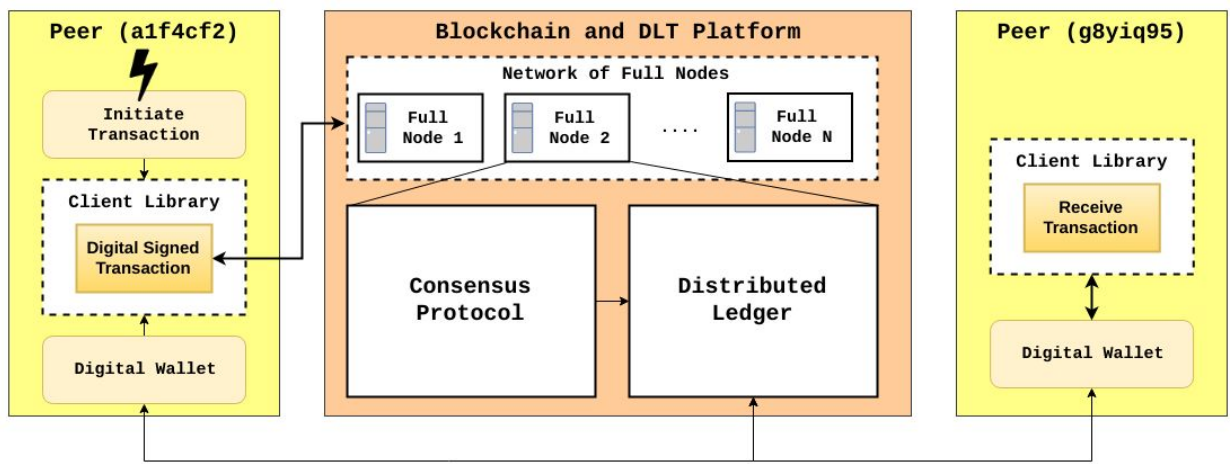

Figure 2: Peer-to-Peer Transaction Using Blockchain Technology.

**Decentralized Computation:** Applications built on top of a blockchain or distributed ledger technology are called "decentralized applications" or Dapps. Such application logic may be executed by the full nodes in the blockchain network by executing smart contracts (e.g., in Ethereum) or through off-chain interactions with computational servers (e.g. in Hyperledger Fabric) following a protocol defined by the underlying blockchain platform.

Figure 2 shows how peer-to-peer transaction is handled through the blockchain technology and the role of cryptography, consensus protocol, and distributed ledger. All blockchain and distributed ledger technology platform follow the interaction pattern described in Figure 2. A peer in a blockchain architecture refers to a device that can initiate a transaction. In contemporary blockchain systems, mobile phones, class-2 devices, and computers act as a peer since the devices require support for cryptography to digitally sign the transactions. Resource-constrained Class-0 and Class-1 devices are not capable of initiating transactions due to their lack of hardware and processing support for encryption and decryption protocols used in blockchain platforms, unless a special hardware is integrated to these devices to run encryption protocols.

Blockchain or DLT platform typically consists of a collection of full nodes, which run the consensus algorithm and maintains the distributed ledger for storing the transactions. Due to the resource-requirements and the networking demands of the full nodes, a server-grade computer is typically required for consensus process, with the exception of Tendermint, which can run even on Raspberry Pi [8]. All the peer-to-peer interactions happen via the blockchain using the client library provided by the underlying platform.

# 4. Reference Architecture for Blockchain-based Peer-to-Peer IoT Applications

The interconnection of IoT deployments or devices to the blockchain technology enable multiple stakeholders to operate in a trustworthy environment through decentralized solutions for identity management, access control, or payment. In this section, we present a reference architecture for developing peer-to-peer and scalable blockchain-based IoT applications.

## 4.1 Interfacing Blockchain Platforms with IoT Deployments

We assume that IoT deployments interact with a blockchain platform to either store or retrieve data from the ledger. When storing data to the blockchain platform, the IoT application or devices may register its identifier, rate the other stakeholders, or announce its willingness to be part of a decentralized application running on the blockchain platform. Similarly, the devices may query the blockchain platform to check the account balance in his wallet or to verify the identity of another device before doing a transaction. Irrespective of the application functionality, IoT applications or devices participate in either writing to and reading from a blockchain using a client library provided by the underlying blockchain platform. An identity management and data monetization applications are used to illustrate the commonly used interaction patterns in peer-to-peer blockchain-based IoT applications:

**Blockchain-based Identity Management Platform for IoT:** Assuming that an identity-management platform manages the identities of all the devices, gateways, edge and cloud service providers through a reliable blockchain-based decentralized application, all the devices register themselves to the platform. As the members in the identity management ecosystem start to use the platform, their reputation increases through a rating mechanism. The rating enables the ecosystem members to rate each other whenever they execute a transaction. When integrating the identity management platform to an IoT application, devices and service providers register themselves using a write operation and query the platform to verify the identity and reputation of users through a read operation.

**Data Monetizations through Micropayments:** Support for micropayments is desired as the IoT device owners start to monetize their IoT data. Each device maintains a digital wallet to buy data or other services from other devices and service providers. On the other side, devices providing services receive money from other digital wallets. In this type of applications, devices initiate a transaction by digitally signing the transaction as shown in Figure 2. The blockchain platform handles the transaction and updates the digital wallet on both the selling and buying peers (see Figure 2). Notably, the transactions are submitted to the blockchain platform using a write operation, while the digital wallet updates and reads its current account balance through a read operation.

Almost all the blockchain-based IoT applications would employ the communication patterns presented in the above application examples. Thus, the architecture for a blockchain-based peer to peer IoT application has to have various communication channels between peers to deliver the desired functionality. Figure 3 presents the key communication channels that we believe are required for building peer-to-peer blockchain-based IoT applications with support for data monetization. The communication channels are discussed below:

**App-specific Transaction Channel:** IoT applications employing blockchain for micropayments and other types of applications require a channel to either write to or read from the blockchain platform. This channel is typically used to interact with a smart contract or other decentralized application hosted on a blockchain platform.

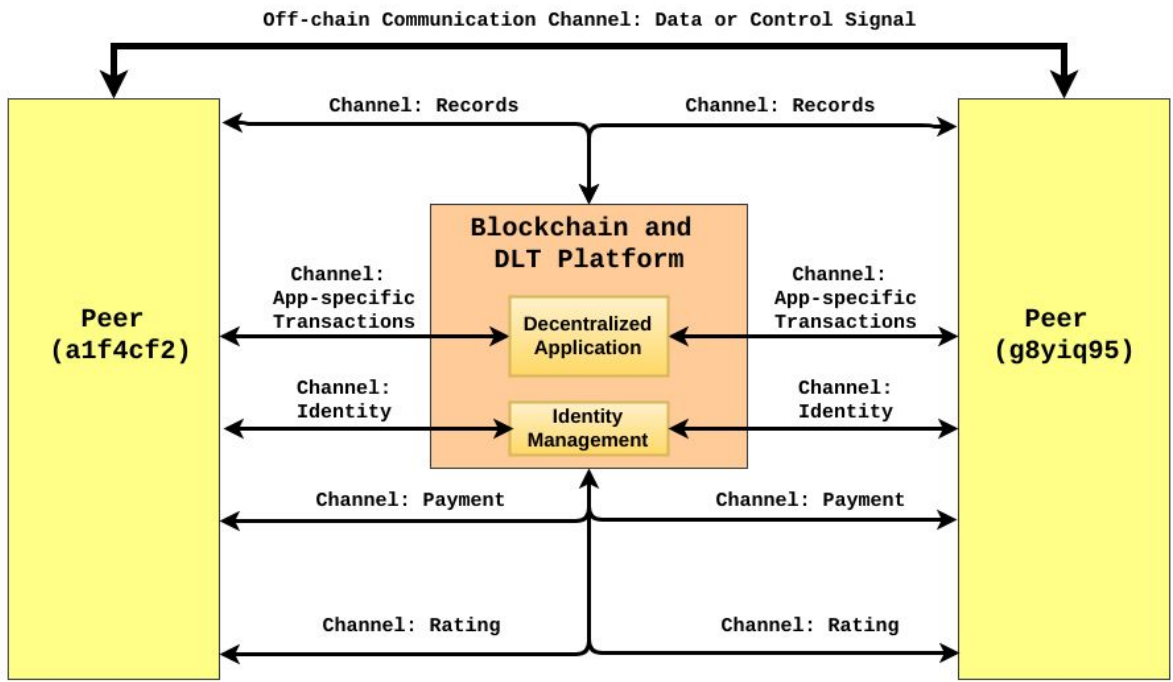

Figure 3: Channels in Our Architecture for Blockchain-based Peer-to-Peer IoT Applications.

**Payment Channel:** This channel is responsible for handling micropayments between peers. The peer that initiates the transaction requires hardware support to prepare and submit the transaction to the blockchain platform. Both the sender and the receiver interacts with the blockchain platform through the payment channel to send and receive money. Peers are expected to maintain a digital wallet to store and manage currencies.

**Identity Channel:** The large-scale IoT deployments are expected to involve multiple stakeholders to perform sensing, processing, and communication functionalities. When IoT devices buy or consume services from other stakeholders, it is important to verify the identity of the service providers and sellers. This channel is used to register, manager, and query identities from the underlying blockchain-based identity management platform.

**Identity Channel:** Peers are executing transactions with other stakeholders in a decentralized environment. This channel is used to manage the identity of the devices, which is a critical component for managing the reputation through a rating mechanism.

**Rating Channel:** Blockchain and DLT technologies allow devices and service providers to provide and consume data and other services from various stakeholders. To make sure that the stakeholders are honest and trustworthy, it is important to maintain the trust and reputation rating of the peers. Misbehaving peers are removed from the platform when their rating is not satisfactory. The Rating channel assumes the availability of identities.

**Records Channel:** This channel records all the transactions in the ledger for future verification. Since the peers exchanging services with each other through a blockchain platform, each and every transaction has to be recorded in the ledger to go back and verify the activities in case of disputes. Note the seller of a data or service may be cheated when the buyer leaves from the application network without paying for the service after consuming a service or data from the buyer. At the other side, the buyer can pay for the service or data prior to consumption and the seller may leave the application network without providing the service or data to the buyer. This problem is referred as "seller's and buyer's dilemma [9]". To prevent disputes in such scenarios, it is important to record all activities such as ordering, invoices, payment receipts, etc. on the blockchain/DLT platform.

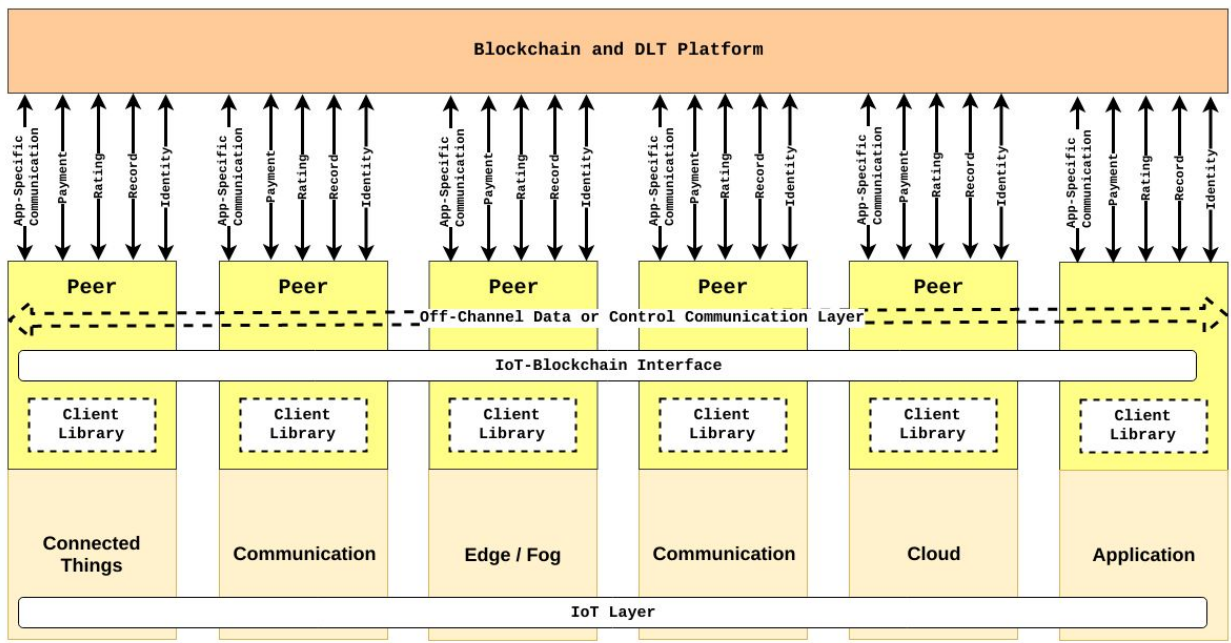

Figure 4: Reference Architecture for Blockchain-based IoT Applications.

**Off-chain Communication Channel:** Devices (or peers) exchanging data or compute services or sending control signals for actuations are not recommended to use the blockchain platform due to the large amount of data. Blockchain platforms are both expensive and less responsive when exchanging large quantity of data. We recommend the application developers to use the blockchain platform only to store the record of data transactions and use a dedicated data channel off-chain through UDP, TCP, or other types of communication frameworks using this communication channel.

Figure 4 shows our proposed reference architecture for blockchain-based IoT applications, showing how devices at IoT layer can interact with the blockchain through a client library using the different communication channels presented in Figure 3. Among all the layers presented in the IoT layer, "connected things" devices are less suitable for running blockchain-based applications due to the resource constraints. Notable exceptions are smartphones and Raspberry-Pi class devices as they possess sufficient resources for digitally signing blockchain transactions using public-key cryptography. Internet connectivity is critical for applications that use public blockchain platforms or even for private blockchain platforms with multiple full nodes.

## 4.2 A Note on Smart Contracts

Some blockchain platforms come with support for smart contracts. IoT applications employing smart contract would still use a client library to digitally sign the transaction before submitting it to a smart contract and hence the interaction of the IoT client with the blockchain would remain the same whether or not a smart contract is used. Functionally, the smart contract provides a way to automate decentralized application functionalities through transaction events, which means each transaction event triggers one or more functions on the smart contract. While our reference architecture is minimalist in that it doesn't require the blockchain platform to provide smart contract functionality (e.g., currently IOTA does not), some peer to peer IoT applications may leverage this additional functionality, as we see in the example of DDM (see Section 6.2) where smart contracts were used for the rating mechanism. Interactions with smart contracts in these cases would utilize the app-specific communication channel.

## 4.3 Guidelines for Application Developers

The reference architecture presented in Figure 4 does not depend on a blockchain platform. Thus, the application developers can select any blockchain platform for their peer-to-peer IoT application. In this section, we provide some additional guidelines for application developers.

**Permissioned vs Public Blockchain Platform:** The choice between permissioned and public blockchain determines the transaction cost and the scalability of the blockchain-based IoT application. On one hand, the public blockchain platforms incurs a small transaction fee whenever a new transaction is submitted plus the slowness of the PoW-type consensus process make the peers wait between few seconds and few minutes for the verification and the storage of transaction in the ledger. Hence, they are not suitable for time-sensitive IoT applications. On the other hand, permissioned platforms are deployed and managed by  stakeholders in application ecosystem, hence such platforms typically employ a lightweight consensus protocol (e.g. PBFT or similar variants) to achieve faster consensus on the transactions. Besides, permissioned platforms such as Hyperledger Fabric do not charge transaction fees unlike public blockchain platforms such as Ethereum and BitCoin. Developers must select the right technology taking into account their application requirements and the pros and cons of different blockchain platforms.

**Cryptocurrency vs Fiat currency:** The reference architecture presented in Figure 4 uses a blockchain platform to enable micropayments between peers. However, the application developers can use cryptocurrency (e.g. ETH and IOTA) or fiat currencies (e.g. USD and EUR) in the payment channel since governmental regulations in certain countries does not permit the use of cryptocurrencies for micropayments.

**Applications with Multiple Blockchain Platforms:** The communication channels presented in Figure 3 illustrate the purpose of different channels, but it does not recommend the application developers to choose a single blockchain platform for payment, record, or rating functionalities. A single peer-to-peer blockchain-based IoT application can employ multiple blockchain platforms to handle application functionalities such as payment and identity management.

# 6. Example Protocols and Frameworks

In this section, we present the frameworks and protocols that uses the reference architecture and the communication channels presented in Figure 4.

## 6.1 Streaming Data Payment Protocol [10]

Streaming data payment protocol (SDPP) enables a buyer and seller to easily connect and transact with each other using micropayments for streaming IoT data. The design carefully separates out three key components: the *off-chain data communication channel* (which is operated as a traditional Internet client-server application-layer protocol, atop TCP), a *payment channel* (implemented using a cryptocurrency protocol), and *a records medium* (implemented using a distributed ledger technology). Figure 5 shows the architecture of SDPP and illustrates how the different communication channels and the reference architecture presented in Figure 4 is used to build a payment solution on top of blockchain and distributed ledger technologies.

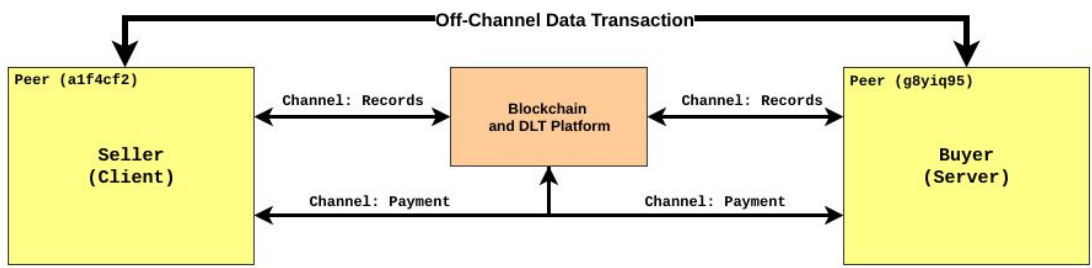

Figure 5: Communication Channels in Streaming Data Payment Protocol.

## 6.2 Decentralized Data Marketplace [11]

Given the economic value associated with IoT data from different parties in the IoT context as shown in Figure 1, it becomes essential for parties involved to sell, find and buy the data as easily as possible. IoT data marketplace platforms are developed for smart cities to address such issues. DDM, a decentralized data marketplace, presents a blockchain-based marketplace for the IoT and smart cities. Architectural elements of data marketplace is distributed through smart contracts and payment channels in DDM. Figure 6 shows how a decentralized data marketplace platform used the reference architecture presented in Figure 5. DDM uses an *application-specific communication channel* to post data products to the marketplace smart contract. Besides, it uses additional channels for *rating and payment transactions*. For exchanging data and services between a seller and a buyer, an *off-chain communication channel* is established between the peers.

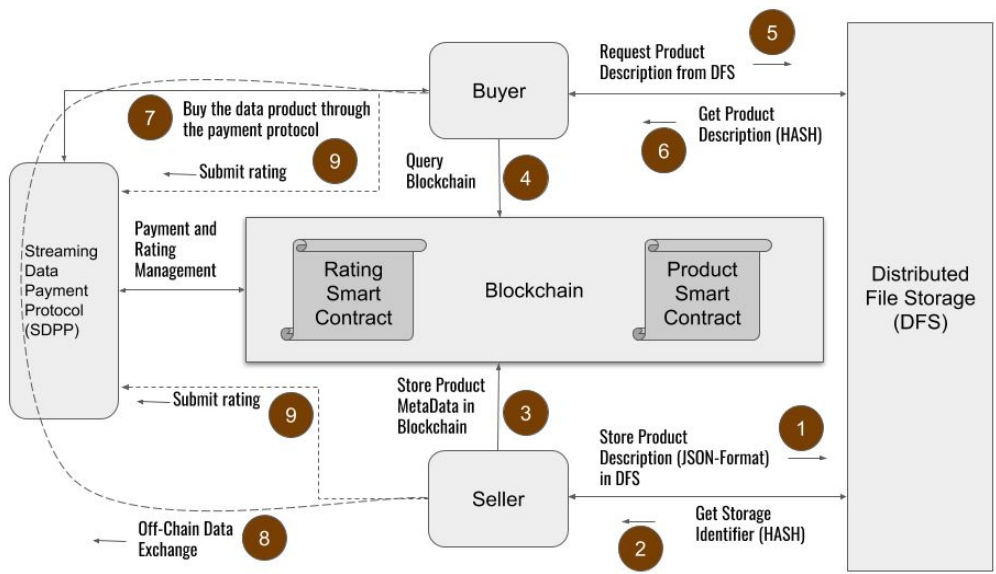

Figure 6: Architecture of Decentralized Data Marketplace.

## 6.3 MOTIVE: Micropayments fOr Trusted vehIcular serVicEs [12]

MOTIVE (an acronym coined from "Micropayments fOr Trusted vehIcular serVicEs"), is a novel vehicle-to-infrastructure and vehicle-to-vehicle (V2x) platform with support for trust management, micropayments, and mechanisms to provide and consume data and compute services with other vehicles and road-side units following a decentralized architecture. MOTIVE is a blockchain and protocol agnostic framework for V2X involving blockchain and distributed ledger technologies, which uses *rating, payment, identity management, and an off-chain communication channel* [12] to reliably exchange data and payment services using a peer-to-peer application model.

## 6.4 Other Examples

FairAccess [13] uses the blockchain technology to create an access control framework for the IoT applications. The implementation of FairAccess consists of a channel for managing and requesting access to remote devices and services through an *application-specific communication channel*. Lin *et al. [14]* presents a framework to store the data from LoRaWAN devices to an immutable storage provided by the blockchain platform by connecting the LoRaWAN gateway to a blockchain platform. This framework uses the *records channel* to store the information on the blockchain. Dorri *et al.* [15] applies the blockchain technology to smart home applications for handling the security and privacy concerns. All these applications connect the IoT applications to the blockchain platform using communication channels that are a subset of those presented in Figure 4. Based on the several examples above, we believe application developers and architects can use our reference architecture to build many other blockchain-based P2P IoT applications.

# 7. Conclusion

The blockchain and distributed ledger technology have the potential to address the scalability, trust, security, and privacy issues of IoT and smart city applications. In large scale IoT applications, the involvement of multiple stakeholders and the emerging data economies introduce the need for new types of architectures and solutions with built-in capabilities for handling trust and payment transactions. In this article, a reference architecture for blockchain-based peer to peer IoT applications has been presented, which highlights the key communication channels needed to interface the IoT layer to the blockchain layer. The usefulness of the proposed architecture has been explained through a systems and protocols developed for blockchain-based peer-to-peer IoT applications.